\documentclass[11pt,twoside]{article}

\usepackage{asp2014}

\aspSuppressVolSlug
\resetcounters

\begin{document}

\title{The Astrophysics Source Code Library: What's new, what's coming}

\author{Alice Allen,$^{1,9}$ G. Bruce Berriman,$^2$ Kimberly DuPrie,$^{3,1}$ Jessica Mink,$^4$ Robert Nemiroff,$^5$ P. Wesley Ryan,$^1$ Judy Schmidt,$^1$ Lior Shamir,$^6$ Keith Shortridge,$^7$ Mark Taylor,$^8$ Peter Teuben,$^9$ John Wallin$^{10}$ and Rein H. Warmels$^{11}$}
\affil{$^1$Astrophysics Source Code Library, College Park, MD, US; \email{aallen@ascl.net}}
\affil{$^2$Caltech/IPAC-NExScI, Pasadena, CA, US}
\affil{$^3$Space Telescope Science Institute, Baltimore, MD, US}
\affil{$^4$Smithsonian Astrophysical Observatory, Cambridge, MA, US}
\affil{$^5$Michigan Technological University, Houghton, MI, US}
\affil{$^6$Lawrence Technological University, Southfield, MI, US}
\affil{$^7$Knave and Varlet, McMahons Point, NSW, Australia}
\affil{$^8$H.~H.~Wills Physics Laboratory, University of Bristol, U.K.}
\affil{$^9$Astronomy Department, University of Maryland, College Park, MD, US}
\affil{$^{10}$Middle Tennessee State University, Murfreesboro, TN, US}
\affil{$^{11}$European Southern Observatory, Garching, DE}

\paperauthor{Alice Allen}{aallen@ascl.net}{0000-0003-3477-2845}{Astrophysics Source Code Library}{}{College Park}{MD}{}{US}
\paperauthor{G. Bruce Berriman}{gbb@ipac.caltech.edu}{orcid.org/0000-0001-8388-534X}{California Institute of Technology}{IPAC-NExScI}{Pasadena}{CA}{}{US}
\paperauthor{Kimberly DuPrie}{kduprie@stsci.edu}{}{Space Telescope Science Institute/ASCL}{}{Baltimore}{MD}{}{US}
\paperauthor{Jessica Mink}{jmink@cfa.harvard.edu}{orcid.org/0000-0003-3594-1823}{Smithsonian Astrophysical Observatory}{}{Cambridge}{MA}{}{US}
\paperauthor{Robert Nemiroff}{nemiroff@mtu.edu}{orcid.org/0000-0002-4505-6599}{Michigan Technological University}{}{Houghton}{MI}{}{US}
\paperauthor{Judy Schmidt}{gecko@geckzilla.com}{orcid.org/0000-0002-2617-5517}{Astrophysics Source Code Library}{}{College Park}{MD}{}{US}\paperauthor{P. Wesley Ryan}{wes@ascl.net}{}{Astrophysics Source Code Library}{}{College Park}{MD}{}{US}
\paperauthor{Lior Shamir}{lshamir@ltu.edu}{orcid.org/0000-0002-6207-1491}{Lawrence Technical University}{}{Southfield}{MI}{}{US}
\paperauthor{Keith Shortridge}{keithshortridge@gmail.com}{orcid.org/0000-0003-1480-9217}{Knave and Varlet}{}{McMahons Point}{NSW}{}{Australia}
\paperauthor{M.~B.~Taylor}{m.b.taylor@bristol.ac.uk}{0000-0002-4209-1479}{University of Bristol}{School of Physics}{Bristol}{Bristol}{BS8 1TL}{U.K.}
\paperauthor{Peter Teuben}{teuben@astro.umd.edu}{orcid.org/0000-0003-1774-3436}{University of Maryland}{Astronomy Department}{College Park}{MD}{}{US}
\paperauthor{Rein Warmels}{rwarmels@eso.org}{0000-0002-9814-0305}{European Southern Observatory}{}{Garching}{}{}{Germany}

\begin{abstract}
The Astrophysics Source Code Library (ASCL, \href{<http://www.ascl.net>}{ascl.net}), established in 1999, is a citable online registry of source codes used in research that are available for download; the ASCL's main purpose is to improve the transparency, reproducibility, and falsifiability of research. In 2017, improvements to the resource included real-time data backup for submissions and newly-published entries, improved cross-matching of research papers with software entries in ADS, and expansion of preferred citation information for the software in the ASCL. 
\end{abstract}


\section{Introduction}
The Astrophysics Source Code Library (ASCL) registers, and in some cases houses, software used in refereed astrophysics research. Its purpose is to improve the transparency, falsifiability, and replicability of the discipline by making the source codes used to produce research results more discoverable. Established in 1999, it is indexed by NASA's Astrophysics Data Service (ADS)\footnote{\url{http://www.adsabs.harvard.edu/}} and Web of Science's Data Citation Index\footnote{\url{http://wokinfo.com/products_tools/multidisciplinary/dci/}} and has inspired the creation of a similar site, the Remote Sensing Code Library\footnote{\url{https://rscl-grss.org/index.php}} \citep{Ulaby2017}. The value of the ASCL is recognized by researchers and indexers in other sciences, and those working on the ASCL have been involved in cross-disciplinary efforts such as Workshop on Sustainable Software for Science: Practice and Experiences \citep{wssspe4}, Force11 Software Citation Implementation Working Group \citep{softwarecitationprinciples}, and CodeMeta.\footnote{\url{https://codemeta.github.io/}}

\section{Growth and activity in 2017}
Use of the ASCL continues to grow. Figure \ref{ASCL_fig1} shows that citations to ASCL entries have increased dramatically over time, with 66\% growth in 2017 (as of November 24) over 2016, and over 60 journals indexed by ADS showing citations to ASCL entries. Submissions by authors have also increased, with the past 12 months showing an increase of 40\%, and the number of entries is nearly 1600. Views of ASCL entries on ADS are up an average of 44\% over 2016, indicating increased discoverability of software through that resource.

 \articlefigure[width=.75\textwidth]{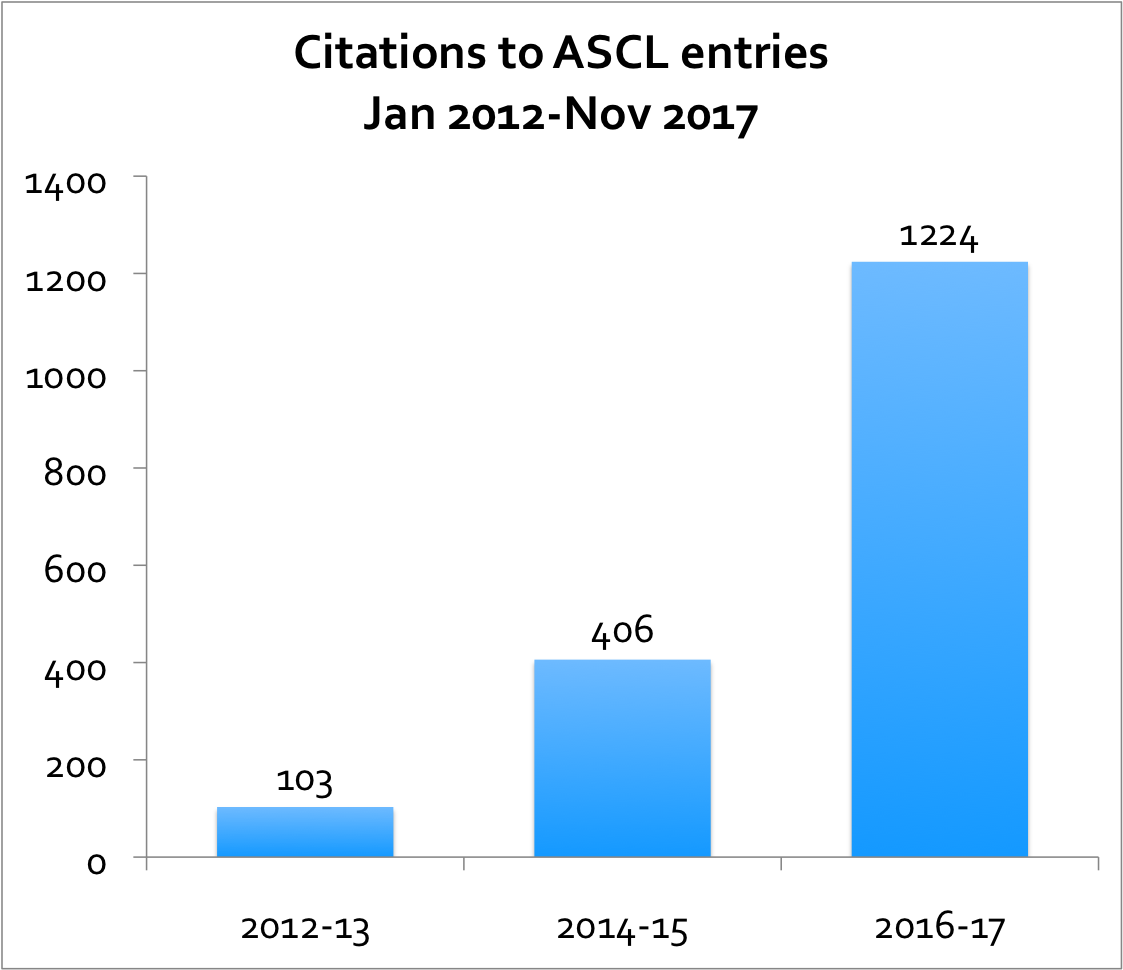}{ASCL_fig1}{Citations to ASCL entries, Jan 2012-Nov 2017} 


In addition to ADASS, the ASCL participated in several other conferences and meetings in 2017. Chief among these were a poster presentation about the ASCL and a Special Session on \textit{Perspectives in Research Software: Education, Funding, Reproducibility, Citation, and Impact}\footnote{http://ascl.net/wordpress/2016/12/06/perspectives-in-research-software-special-session-at-aas-229/}, organized by the ASCL and the Moore-Sloan Data Science Environment at NYU, at the January American Astronomical Society (AAS) meeting in Texas and the European Week of Astronomy and Space Science (EWASS) in June, held in Prague, for which the ASCL partnered with others in presenting a Special Session on \textit{Developments and Practices in Astronomy Research Software}.\footnote{http://ascl.net/wordpress/2017/07/23/special-session-on-and-about-software-at-ewass-2017/} 

The ASCL also participated in the first ever EWASS Hack Day. Three Hack Day participants worked with the ASCL editor on two different projects, one to create metadata for the as-yet-unused category field in the ASCL by mining papers to which the ASCL links for keywords and associating them with ASCL entries, and another to add preferred citation information to entries where it was missing. This latter project was so successful that not only were 52 entries edited at or shortly after EWASS to add this information, one participant continued to add data to the online file used to gather the information in the months after EWASS. Anonymous others also provided information, as the link to the data collection site was tweeted out during and after the Hack Day.

\section{Improvements}
Early in 2017, the ASCL's emergency move from one server to another made us realize that we could do a better job in preventing data loss. No data were lost in the move, thanks to the excellence of the MTU technical team, but we saw a potential for the loss of brand new submissions to the ASCL should a problem develop in the hours between their submission and the daily backup of the ASCL. New submissions already generated an email to members of the ASCL team with some metadata for the submission, but did not include all of it. The auto-email system was made more robust by not only including all submitted metadata in the submissions-generated email, but also by capturing this information off-site in a secure location immediately upon submission of the entry.

We have improved the cross-matching of software with articles in ADS by replacing links to articles in resources such as arXiv with links to ADS entries, which are used to create the cross-matching (example shown in Figure \ref{ASCL_fig2}) and continue to do so; we expect to complete this work by the end of 2017. This will increase the discoverability of codes by making it possible to find them from ADS entries for articles not only via the citations list, but also from the article's abstract page. 

\articlefigure[width=.95\textwidth]{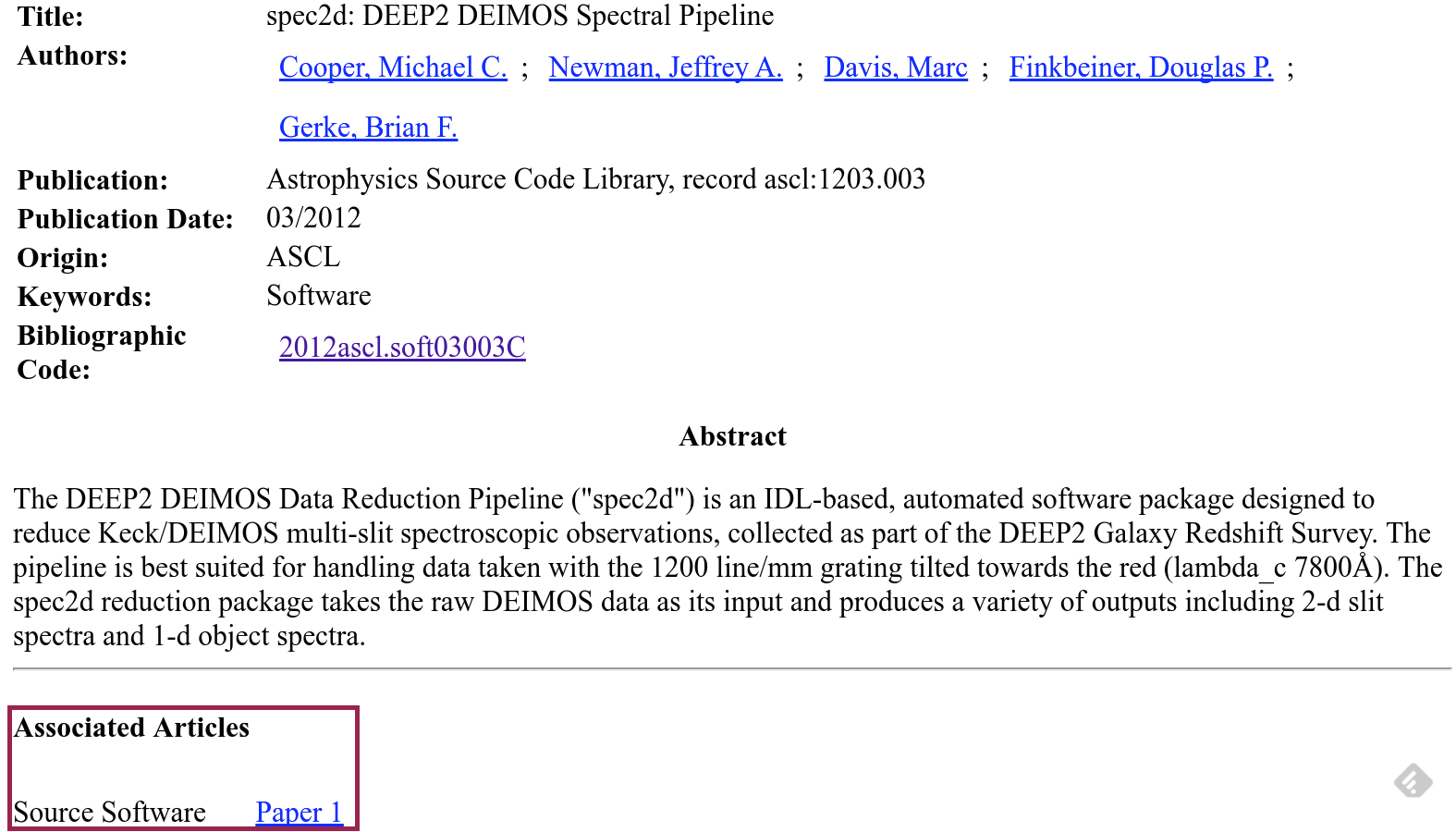}{ASCL_fig2}{Screenshot of ASCL entry in ADS showing Associated Articles information} 


\section{Future plans}
The ASCL plans to expand the metadata it contains and to improve its search capabilities. The ASCL is partnering with others to organize sessions at future meetings, including a Special Session at the January 2018 AAS meeting and a Symposium, which is two days of talks, at the 2018 joint Royal Astronomical Society Annual Meeting/EWASS that will be held in April in Liverpool. We hope to do additional outreach in 2018 to universities and organizations to inform them of the ASCL, and will continue to meet with and discuss common issues with those who maintain similar resources in other disciplines.

\acknowledgements The ASCL is grateful for financial support provided by the Heidelberg Institute for Theoretical Studies (HITS) and for facilities and services support from Michigan Technological University, the Astronomy Department at the University of Maryland, and the University of Maryland Libraries.

\bibliography{P11-151}  

\end{document}